\documentstyle[a4wide,12pt,epsf]{article}
\begin{document}
% TH FORMAT
\begin{flushright}
\def\SLZ{{$SL(2,\IZ)$}}
\def\IZ{\relax\ifmmode\mathchoice
{\hbox{\cmss Z\kern-.4em Z}}{\hbox{\cmss Z\kern-.4em Z}}
{\lower.9pt\hbox{\cmsss Z\kern-.4em Z}}
{\lower1.2pt\hbox{\cmsss Z\kern-.4em Z}}\else{\cmss Z\kern-.4em
Z}\fi}

\baselineskip=12pt
{SUSX-TH-98-008}\\
{hep-th/9808142}\\
{August 1998}
\end{flushright}

\begin{center}
%\vglue 0.5cm
{\LARGE \bf Cosmological inflation with orbifold moduli as inflatons \\}
\vglue 0.35cm
{D.BAILIN$^{\clubsuit}$ \footnote
{D.Bailin@sussex.ac.uk}, G. V. KRANIOTIS$^{\spadesuit}$ \footnote
 {G.Kraniotis@rhbnc.ac.uk} and A. LOVE$^{\spadesuit}$ \\}
%\vglue 0.2cm
	{$\clubsuit$ \it  Centre for Theoretical Physics, \\}
{\it University of Sussex,\\}
{\it Brighton BN1 9QJ, U.K. \\}
%\vgluw 0.2cm
{$\spadesuit$ \it  Department of Physics, \\}
{\it Royal Holloway and Bedford New College, \\}
{\it  University of London,Egham, \\}
{\it Surrey TW20-0EX, U.K. \\}
\baselineskip=12pt

\vglue 0.25cm
ABSTRACT
\end{center}

%\vglue 0.5cm
{\rightskip=3pc
\leftskip=3pc
\noindent
\baselineskip=20pt
Cosmological inflation is studied in the case where the inflaton is the 
overall modulus $T$ for an orbifold. General forms 
of the (non-perturbative) superpotential are considered  to 
ensure that $G=K+{\rm ln}|W|^2$ is modular invariant.  
We find generically
that these models do not produce a potential flat enough for slow 
roll to a supersymmetric minimum, although we do find a model which 
produces up to 20 e-folds of inflation to a non-supersymmetric 
minimum.}

\vfill\eject
\setcounter{page}{1}
\pagestyle{plain}
\baselineskip=14pt

The need for an inflationary stage \cite{GUTH} in the early universe is 
well known. (For reviews see ref.\cite{OLIVE}) Cosmological 
inflation can solve the 
flatness and horizon problems and, because of de Sitter fluctuations during 
inflation, may also account for the observed density and temperature 
fluctuations. In the new inflationary universe scenario \cite{LINDE,PAUL} 
the expectation value of a scalar field (the inflaton) rolls slowly in a 
region of the effective potential which is extremely flat with a 
positive vacuum energy during inflation. Eventually, it evolves more 
rapidly and oscillates about the minimum of the effective potential 
converting the vacuum energy into radiation and reheating the universe.

Good candidates \cite{PIERRE,Freese} for the role of inflaton are the 
dilaton and moduli fields in orbifold or Calabi-Yau compactifications 
of string theory. To all orders in string perturbation theory their 
potential is completely flat \cite{DKL}. Non-perturbative effects, 
such as gaugino condensation, can provide a non-trivial effective 
potential with variation on the Planck scale as the dilaton or 
modulus expectation value varies. Even then, the generic result to be 
expected \cite{MOORE} is that the number of $e$-folds of inflation 
$N_e$ will be of order 1 rather than $N_e\sim 60$ as required to 
succeed in solving the various cosmological problems.

There have been some attempts to employ the dilaton $S$ as inflaton 
\cite{PIERRE,Freese}. Unfortunately, dilaton dynamics is little 
understood and, consequently, the conclusions reached have been 
largely qualitative. However, if the relevant dynamics is a multiple 
gaugino condensate \cite{Freese} that fixes the dilaton expectation 
value at the minimum at a realistic value with regard to the 
strength of the gauge coupling constant, the effective potential may 
be sufficiently flat in the $Im S$ direction. On the other hand, the 
potential would be expected to very steep in the $Re S$ direction 
\cite{BRUSTEIN}. 

Instead, we shall focus here on orbifold moduli as candidate inflatons. 
The reason for doing this is that moduli 
dynamics is more constrained because it is subject to modular symmetries 
which are believed to be symmetries of the theory not only to all 
orders in string perturbation theory but also when non-perturbative 
effects are included (see, for example, ref.\cite{MIRIAM} 
and references therein.) We study orbifolds rather than Calabi-Yau 
manifolds because the form of the moduli K$\rm{\ddot a}$hler potential 
is not known in the Calabi-Yau case, except for large values of the 
moduli $T_i$, because is difficult to calculate world-sheet instanton 
contributions \cite{CANDELAS}. We shall focus in the case of a single 
overall modulus $T_1=T_2=T_3=T$. Unlike some other authors 
\cite{COPELAND} we treat the modulus field itself as the inflaton 
rather than some combination that involves a matter field scalar and 
also work in the context of the standard new inflationary scenario rather 
than in the context of hybrid inflation \cite{ALINDE,HITO}.

It will be assumed that the effective superpotential is the sum 
of two components \cite{MOORE}. One of these components 
has a large scale, and gives
an effective potential with unbroken supersymmetry 
and zero cosmological constant at the minimum when the other component 
is neglected. The large scale component is responsible for driving 
inflation when the modulus expectation value is in a flat region 
away from the minimum. The other component has a much smaller scale and 
is responsible for the supersymmetry breaking in the low energy world.
It is the former component of the superpotential that we are interested in 
here. Neglecting the low energy component of the superpotential, it is 
convenient to write the effective potential $V$ in the form 
\begin{equation}
V=\mu^4 F(T/\tilde{M}_p)
\label{V}
\end{equation}
where $F$ is of order 1,
\begin{equation}
\tilde{M}_p \approx 2.44 \times 10^{18} \rm{GeV}
\label{PLANCK}
\end{equation}
and \cite{THOMAS}
\begin{equation}
\mu \sim 10^{16} \rm{GeV}
\label{scale}
\end{equation}
in order to obtain density perturbations $\frac{\delta \rho}{\rho}\sim
5 \times 10^{-5}$ as required by COBE data. This contrasts with a scale 
of $10^{10}-10^{11} \rm{GeV}$ if the component of the effective potential 
were due to the component of the $W$ responsible for low energy 
supersymmetry breaking with soft supersymmetry breaking masses in the 
scale of $10^2-10^3 \rm{GeV}$. At the end of inflation, we need the 
modulus $T$ to have reached a supersymmetric minimum of the effective 
potential to avoid large supersymmetry breaking in the low energy theory and 
to avoid a large cosmological constant in the present universe. Thus,  we 
require a model in which there is at least a local supersymmetric minimum 
of the effective potential with zero cosmological constant. We also 
require that the effective potential should have a maximum or saddle point 
in the vicinity of which the potential is sufficiently flat for 
$60$ e-folds of inflation to occur. We shall come back to possible models 
after we have set out the conditions for slow roll and an estimate 
for $N_e$ in terms of the effective potential and its derivatives.

The discussion of slow roll for the modulus field $T$ differs from the 
standard case \cite{LINDE,PAUL,TURNER} in two respects. First, $T$ is 
a complex scalar field rather than a real scalar and rolls in a 2-dimensional 
space. Second, $T$ has a non-minimal K$\rm{\ddot a}$hler potential and 
corresponding non-minimal kinetic terms. (For reviews of 
supergravity and superstrings see ref. \cite{DBAL} )
The tree level K$\rm{\ddot a}$hler potential for the overall orbifold 
modulus $T$ takes the form
\begin{equation}
K=-3 {\rm ln} (T+\bar{T})
\label{tree}
\end{equation}
We shall allow for the possibility of stringy non-perturbative 
corrections \cite{SHENKER} and write 
\begin{equation}
K=Q(T+\bar{T})
\end{equation}
Then, the Lagrangian for $T$ is 
\begin{equation}
{\cal L}=\frac{\partial^2 Q}{\partial T \partial{\bar{T}}}
\partial_{\mu}{T}\partial^{\mu}{\bar{T}}-V
\end{equation}
Assuming a homogeneous field $T$ with zero spatial gradients,
the covariantized field equations are
\begin{equation}
\frac{\partial^3 Q}{
\partial T^2 \partial {\bar{T}}}\dot{T}^2+
\frac{\partial V}{\partial \bar{T}}+
\frac{\partial^2 {Q}}{\partial T\partial {\bar{T}}}(
\ddot{T}+3 H \dot{T})=0
\end{equation}
where $H$ is the Hubble constant given by
\begin{equation}
H^2=\frac{\rho}{3}=\frac{1}{3}\frac{\partial^2 Q}{\partial T 
\partial {\bar{T}}}\dot{T}\dot{\bar {T}}+\frac{V}{3}
\end{equation}
in units where 
\begin{equation}
{\kappa }^2=8\pi G_N=\tilde{M}_p^{-2}
\end{equation}
has been taken to be 1, and neglecting the curvature of the universe. 
Writing $T$ in terms of its real and imaginary parts 
\begin{equation}
T=T_1+i\; T_2
\end{equation}
the slow roll equations are
\begin{equation}
6 H \frac{\partial^2 Q}{\partial T 
\partial {\bar{T}}}\dot{T}_1=-\frac{\partial V}{\partial T_1}
\end{equation}
and 
\begin{equation}
6 H \frac{\partial^2 Q}{\partial T 
\partial {\bar{T}}}\dot{T}_2=-\frac{\partial V}{\partial T_2}
\end{equation}

If the energy density is dominated by the potential energy, so that 
$H^2 \approx \frac{V}{3}$, as we shall assume in what follows, then 
we require
\begin{equation}
\frac{\Bigl(\frac {\partial^2 Q}{\partial T 
\partial {\bar{T}}}\Bigr)^{-1}}{12}
\frac{\Bigl( (\frac{\partial V}{\partial{T_1}})^2+
(\frac{\partial V}{\partial{T_2}})^2 \Bigr)}{V^2}\ll 1
\label{density}
\end{equation}

For the slow roll approximation to be valid it is necessary to have 
\begin{equation}
\Biggl|\frac{\ddot {T}_1}{3 H \dot{T}_1}\Biggr |\;,\;
\Biggl|\frac{\ddot {T}_2
}{3 H \dot{T}_2}\Biggr |\ll 1
\end{equation}
which can be cast as the sufficient conditions
\begin{eqnarray}
\frac{1}{6}& |q|^{-2}&\Bigl|V^{-1} 
\frac{\partial V}{\partial {T_1}}(\frac{\partial q}{\partial T}+
\frac{\partial q}{\partial{\bar{T}}})\Bigr|\ll 1, \nonumber \\
\frac{1}{6}& |q|^{-2}&\Bigl|V^{-1}
\frac{\partial V}{\partial {T_2}}(\frac{\partial q}{\partial T}-
\frac{\partial q}{\partial{\bar{T}}})\Bigr|\ll 1, \nonumber \\
\frac{1}{6}& |q|^{-1}&\Bigl|(V\frac{\partial V}{\partial {T_1}})^{-1}
\Bigl(\frac{{\partial^2} V}{\partial {T_1}^2}
\frac{\partial V}{\partial {T_1}}+\frac{\partial V}{\partial {T_2}}
\frac{{\partial^2} V}{\partial {T_1}\partial {T_2}}\Bigr)\Bigr|\ll 1, 
\nonumber \\
\frac{1}{6}& |q|^{-1}&\Bigl|(V\frac{\partial V}{\partial {T_2}})^{-1}
\Bigl(\frac{{\partial^2} V}{\partial {T_2}^2}
\frac{\partial V}{\partial {T_2}}+\frac{\partial V}{\partial {T_1}}
\frac{{\partial^2} V}{\partial {T_1}\partial {T_2}}\Bigr)\Bigr|\ll 1
\end{eqnarray}
where we have written
\begin{equation}
q\equiv \frac{\partial^2 Q}{\partial T \partial {\bar{T}}}
\end{equation}
and have dropped a condition identical to (\ref{density})
It is also necessary to have
\begin{equation}
\frac{1}{3H}\Biggl|\frac{\frac{\partial q}{\partial T}}{q}\dot{T_1}\Biggr|,
\;\;
\frac{1}{3H}\Biggl|\frac{\frac{\partial q}{\partial T}}{q}
\dot{T_2}\Biggr|\ll 1
\end{equation}
which when the slow roll is consistent can be cast as
\begin{equation}
\frac{1}{6}\Biggl|q^{-2}\frac{\partial q}{\partial T}V^{-1}
\frac{\partial V}{\partial {T_1}}\Biggr|\;,\;
\frac{1}{6}\Biggl|q^{-2}\frac{\partial q}{\partial T}V^{-1}
\frac{\partial V}{\partial {T_2}}\Biggr|\ll 1
\end{equation}

If the slow roll starts from a point $T_1=(T_1)_{0}\;,\;T_2=(T_2)_0$ 
close to a saddle point or maximum of the effective potential then 
we may write
\begin{eqnarray}
\frac{\partial V}{\partial {T_1}}&\approx & {\Bigl(
\frac{\partial V}{\partial {T_1}}\Bigr)}_{0}+
{\Bigl(\frac{{\partial^2} V}{\partial {T_1}^2}\Bigr)}_{0}{(
T_1-(T_1)}_{0})+{\Bigl(\frac{{\partial^2} V}
{\partial {T_1}\partial {T_2}}\Bigr)}_{0}{(T_2-(T_2)}_0), \nonumber \\
\frac{\partial V}{\partial {T_2}}&\approx & {\Bigl(
\frac{\partial V}{\partial {T_2}}\Bigr)}_{0}+
{\Bigl(\frac{{\partial^2} V}
{\partial {T_1}\partial {T_2}}\Bigr)}_{0}{(T_1-(T_1)}_0)+
{\Bigl(\frac{{\partial^2} V}{\partial {T_2}^2}\Bigr)}_{0}{(
T_2-(T_2)}_0)
\end{eqnarray}

Writting
\begin{equation}
X_{\alpha}=T_{\alpha}-{(T_{\alpha})}_0\;,\;\alpha=1,2
\end{equation}
then, correct to linear order in the $X_{\alpha}$, the slow roll 
equations may be written as
\begin{equation}
\dot{X}=A- M X
\end{equation}
with 
\begin{equation}
X=\pmatrix{X_1\cr X_2\cr}\;,\;A=-\frac{q_0^{-1}}{6H}\pmatrix{
{(\frac{\partial V}{\partial {T_1}})}_0 \cr {(
\frac{\partial V}{\partial {T_2}})}_0 \cr},
\end{equation}
\begin{equation}
M=\frac{q_0^{-1}}{6H}\pmatrix{{(\frac{{\partial^2} V}{\partial {T_1
}^2})}_{0} & {(\frac{{\partial^2} V}
{\partial {T_1}\partial {T_2}})}_{0} \cr
 {(\frac{{\partial^2} V}
{\partial {T_1}\partial {T_2}})}_{0} & 
{(\frac{{\partial^2} V}{\partial {T_2}^2})}_{0} \cr}
\end{equation}

After we diagonalizing $M$ we find that the displacements 
$X_{\alpha}$ are superpositions of eigensolutions with time 
dependence $e^{-\lambda_i t}$ where 
\begin{equation}
\lambda_i=\frac{q_0^{-1}}{6H}\mu_i\;,\; i=1,2
\end{equation}
\begin{equation}
2\mu_{1,2}=(N_{11}+N_{22})\pm \sqrt{(N_{11}-N_{22})^2+4 N_{12}^2}
\end{equation}
and
\begin{equation}
N_{11}=\pmatrix{{\frac{{\partial^2} V}{\partial {T_1
}^2}}\cr}_0\;,\; N_{12}=\pmatrix{{\frac{{\partial^2} V}{\partial {T_1}
\partial {T_2}}}\cr}_{0}\;,\;N_{22}=\pmatrix{{\frac{{\partial^2} V}
{\partial {T_2}^2}}\cr}_{0}
\end{equation}

The number of $e$-folds of inflation may then be written as 
\begin{equation}
N_{e}=2 V_0\; q_0\; min\Bigl\{-\mu_1^{-1},-\mu_2^{-1}\Bigr\}
\label{efold}
\end{equation}
if $\mu_1$ and $\mu_2$ are both negative. Otherwise, $N_e$ is 
controlled by the negative $\mu_i$. (Notice that $q_0>0$ for 
correct sign kinetic terms.)

To avoid de Sitter fluctuations driving $T$ across the slow roll 
region more rapidly than the semi-classical motion we require 
\cite{VILENKIN,TURNER}
\begin{equation}
\Biggl|\frac{\Delta T_1}{{(T_1)}_0}\Biggr|\;,\;
\Biggl|\frac{\Delta T_2}{{(T_1)}_0}\Biggr|\geq
\frac{(2 q_0)^{-1/2} N_e^{1/2} V_0^{1/2} \tilde{M_P}^{-2}}{
2\pi (T_1)_0}
\label{fluctuation}
\end{equation}
where $\Delta T_1,\Delta T_2$ quantify the 
width of the slow roll region.
Putting $N_e \sim 60$ and 
demanding $\tilde{M_P}^{-2} V_0^{1/2}\sim 10^{-4}$
for consistency with $\frac{\delta\rho}{\rho}\sim 5 \times 10^{5}$,
(\ref{fluctuation}) requires
\begin{equation}
\Biggl|\frac{\Delta T_1}{{(T_1)}_0}\Biggr|\;,\;
\Biggl|\frac{\Delta T_2}{{(T_1)}_0}\Biggr|\geq
\frac{(2 q_0)^{-1/2}}{(T_1)_0} 10^{-4}
\end{equation}

It is convenient to focus on the value of $N_e$ in what follows.
When a small value of $N_e$ is obtained we would not expect the 
slow roll conditions to be satisfied and the small value of 
$N_e$  is a symptom of this rather than a reliable determination 
of $N_e$. The outcome for the number of $e$-folds of inflation 
depends on the choice of the (non-perturbative) superpotential 
and of the modulus K$\rm{\ddot a}$hler potential. We shall take 
the  superpotential to be of the form
\begin{equation}
W=\Omega(S) \tilde{H}(T) \eta^{-6}(T)
\label{super}
\end{equation}
where
\begin{equation}
\tilde{H}(T)=\Bigl(j(T)-1728\Bigr)^{m/2} j^{n/3}(T) P(j(T))
\label{modular}
\end{equation}
In (\ref{super}), $\eta(T)$ is the Dedekind eta function and we do 
not specify $\Omega(S)$ which depends on the little understood 
dilaton dynamics. In (\ref{modular}), $j(T)$ is the absolute 
modular invariant, $m$ and $n$ are positive integers and $P$ is 
a polynomial. This form of $\tilde{H}(T)$ is the most general 
form consistent with $PSL(2,Z)$ modular invariance and the 
absence of singularities in the fundamental domain of the 
modular group \cite{MIRIAM} when the modulus 
K$\rm{\ddot a}$hler potential given by (\ref{tree}) or transforms 
in the same way.
Taking the dilaton K$\rm{\ddot a}$hler potential to have the 
standard form
\begin{equation}
K=-{\rm ln} (S+\bar{S})
\end{equation}
(neglecting the effects of the small Green-Schwarz coefficient)
the corresponding effective potential $V$ is given by 
\begin{eqnarray}
V&=& (S+\bar{S})^{-1}\Bigl|\Omega(S)\Bigr|^2 e^{Q(T,\bar{T})}
\Bigl|\eta(T)\Bigr|^{-12}\Bigl|\tilde{H}(T)\Bigr|^{2} \nonumber \\
&\times& \Biggl\{|F_S|^2 -3 \nonumber \\
&+&\Bigl(\frac{\partial^2 Q}{\partial T \partial{\bar{T}}}
\Bigr)^{-1}\Biggl|\frac{d {\rm ln} \tilde{H}}{dT}+
\frac{3 \hat{G}_2}{2\pi}+\frac{\partial Q}{\partial T}+
3(T+\bar{T})^{-1}\Biggr|^2\Biggr\}
\end{eqnarray}
where
\begin{equation}
\hat{G}_2(T,\bar{T})=-2\pi(T+\bar{T})^{-1}-4\pi \eta^{-1}\frac{d\eta}{dT}
\end{equation}
and
\begin{equation}
F_S=1-(S+\bar{S})\frac{d{\rm ln}\Omega}{dS}
\end{equation}
We shall take $F_S=0$ so that the dilaton auxiliary field is zero and there 
is a possibility of finding a minimum with unbroken supersymmetry.

The case where the modulus K$\rm{\ddot a}$hler potential is given by 
the tree level form (\ref{tree}) will be considered first. 
When the non-perturbative superpotential is of the minimal form 
given by (\ref{super}) with $m=n=0,\;P(j)=1$, it is well known 
\cite{LUST} that, $V$ has a single minimum with negative 
vacuum energy and broken supersymmetry (see Fig.1).
As discussed earlier, both of the features are undesirable for the 
high energy component of the effective potential. Moreover, 
$V$ is not even positive at the maximum and there is no positive 
vacuum energy to derive inflation. 
However, the potential is flat enough for slow roll to occur.
An expression of this flatness is that values of $N_e$ as large 
as $\sim 2.5$ can be obtained by rolling from close to the 
maximum if (unjustifiably) we replace $V_0$ by $|V_0|$ in 
(\ref{efold}).

The situation is different if at least one of $m$ and $n$ is positive in 
(\ref{modular}), but keeping $P(j)=1$.
A superpotential of this form may perhaps arise in gaugino condensate 
models with certain gauge non-singlet states becoming massless at some special 
value of $T$ \cite{MIRIAM} or through some other non-perturbative mechanism.
Then, it is often possible for the effective potential to possess at least
one minimum with zero vaccum energy and unbroken supersymmetry, though 
such minima are never the absolute minimum. Since we have no reason to 
exclude the possibility that we are in a long-lived metastable vacuum, 
we study the scenario in which the expectation value of $T$ rolls from 
near a saddle point or maximum of the potential to such a minimum. It is 
known \cite{MIRIAM} that, there is a local minimum at $T=e^{i\pi/6}$ with 
$V=0$ and unbroken supersymmetry whenever $n\geq 2$.
It is also known that there is such a minimum at $T=1$ when 
$m\geq 2$. We have investigated the saddle points and maxima of the 
effective potential (with $P(j)=1$) in case where there is a supersymmetric 
minimum with $V=0$, see Figs 2,3,4.  In all of these cases there are 
saddle points of the potential at which $V$ has very different 
(positive) values.
Allowing the expectation value of $T$ to start rolling from close to a 
saddle point or maximum, we can use (\ref{efold}) to calculate the 
number of $e$-folds of inflation (assuming that the slow roll conditions 
are satisfied.) The maximum values of $N_e$ obtained are 
summarized in tables 1-2.
It can be seen that except for 
$m=0,n=1$ in no case is $N_e$ greater than $10^{-2}$.
In these circumstances, the value of $N_e$ is a symptom that the 
potential is not sufficiently flat for slow roll to occur rather than a
reliable determination of the number of $e$-folds of inflation
\footnote{Note though  that in the case 
$m=2,n=0$ (see Tables 1-2),
besides the zero 
energy supersymmetric minimum we have a global negative energy 
minimum with SUSY broken.
Also there is a negative energy 
local maximum around the vicinity of such minima 
at the fixed point  
of moduli space, $T=\frac{\sqrt{3}}{2}+\frac{1}{2}\;i$. 
Interestingly, if we allow $T$ to roll from the neighbourhood of such a point 
the potential is flat enough for slow roll to occur and 
we obtain a number of $e$-folds $N_e$, of order 1 to 10 
if we again (unjustifiably) replace  $V_0$ by $|V_0|$ in 
(\ref{efold}).
 The same discussion applies to case $m=1,n=0$ although in the latter 
case we do not have a supersymmetric minimum at all.}
As discussed earlier, the generic result for an effective potential 
with variation on the Planck scale with respect to the inflaton 
expectation value would be expected to be inflation with values of 
$N_e$ of order 1. The outcome here is non-generic but in the 
direction of smaller amounts of inflation (or no inflation at all) 
rather than in the direction of enhanced inflation. This is 
essentially a consequence of $\frac{j^{''}}{j}$ and $
\Bigl(\frac{j^{'}}{j}\Bigr)^{2}$ being in the range $10^2-10^3$ 
close to a saddle point or maximum.
However, for $m=0,n=1$, i.e $W_{np}=\eta^{-6}j^{1/3}$ we obtain a 
much flatter potential with negative non-
supersymmetric minimum at $T=1.0$. In this case, the 
saddle point is at positive energy (fig.5) and we obtain 
a number of $e$-folds of order 1 to 10. In particular,
if we start rolling from 
$T\sim 0.862+0.505\; i$ we obtain $N_e \sim 20$.  This is an interesting 
example even though it does not satisfy all of our constraints.

The  K$\rm{\ddot a}$hler potential may be modified by 
stringy non-perturbative effects \cite{SHENKER}. In that case,
we have considered some illustrative examples consistent 
with the K$\rm{\ddot a}$hler potential continuing to transform 
in the same way under modular transformations as (\ref{tree}). 
This is required if 
\begin{equation}
G=K+{\rm ln}|W|^2
\end{equation}
with $W$ given by (\ref{super}) and (\ref{modular}), is to remain 
modular invariant. The specific examples we have considered are 
\begin{eqnarray}
K&=& Q(T+\bar{T})=-3 {\rm{ln}} (T+\bar{T})+
 a\; {\rm{ln}}\; (j(T)+\bar{j}(\bar{T})) \\
K&=& Q(T+\bar{T})=-3 {\rm{ln}}(T+\bar{T})+ 
\tilde{a}\; (j(T)+\bar{j}(\bar{T}))^{p} \\
K&=& Q(T+\bar{T})=-3 {\rm{ln}} (T+\bar{T})+\hat{a}\;(T+\bar{T})|\eta(T)|^4
\end{eqnarray}
Similar results were 
obtained as in  the case of minimal K$\rm{\ddot a}$hler potential.

In conclusion, typically models with overall orbifold modulus as the 
inflaton do not produce a potential flat enough for slow roll.
In the case $W_{np}=\eta^{-6} j^{1/3}$ slow roll does occur with 
as much as 20 $e$-folds of inflation.
However, it does not appear possible to 
obtain a model which produces a minimum with $V=0$ and unbroken 
supersymmetry when slow-roll occurs.
Thus we are unable to explain the scale of the observed 
density fluctuations by cosmological inflation which preserves 
supersymmetry. However, 
it has been observed 
that inflation by about 12 e-folds can be  sufficient to solve the 
moduli problem, namely the domination of the energy density by 
massive scalar fields which do not decay before nucleosynthesis; in the 
present context the $T$-moduli acquire masses $m_T\sim m_{3/2}\sim 
1$ TeV when supersymmetry is broken.
To  solve the moduli problem  \cite{LS} the 
potential $V_0$ at the end of inflation
must satisfy $V_0^{1/4}< 10^{7}-10^8$ GeV. In our case,  the potential 
at the supersymmetry-breaking  
minimum satisfies $V_0^{1/4} \sim \sqrt{m_{3/2}m_P}\sim 
10^{11}$ GeV, so the moduli problem too remains unsolved.

\section*{Acknowledgements}
This research is supported in part by PPARC.

\newpage

\begin{table}
\begin{center}
\begin{tabular}{|c|c|c|c|c|}  \hline\hline
$m$ &$n $ & $V_{saddle/max}$ & $V_{min}$  
& $T_{max}/T_{saddle}$ \\ \hline\hline
0   & 0 &-8.41  &-8.93 &$1+\frac{1}{2}\;i/ 1+n\;i,\frac{\sqrt{3}}{2}+
\frac{1}{2}\;i$ \\
1   & 3 & $O(10^{11})$ & 0 & $\sim 1+0.3\;i$ \\
2   & 0 & $O(10^{8})/-O(10^7)$ & $0/
-O(10^7)$ 
& $\frac{\sqrt{3}}{2}+\frac{1}{2}\;i / \sim 1+0.2\;i$ \\
2   & 3 & $O(10^{14})$ & $0/-O(10^{12})$  
&$1.0+0.1\;i,1.0+0.3\;i$ \\
0   & 1 & 3550.97 & -1277.99   
&$\sim O(1)+\frac{1}{2}\;i$\\
\hline\hline
\end{tabular}
\end{center}
\caption{Extrema of $V_{eff}$ for  different choices of the integers $m,n$}
\end{table}

\begin{table}
\begin{center}
\begin{tabular}{|c|c|c|}  \hline\hline
$m$ &$n $ & range of $N_{e}$ 
\\ \hline\hline
0   & 0 & $\sim 2.5$ \\
1   & 3 & $2\times 10^{-3}-5 \times 10^{-3}$ \\
2   & 0 & ${\sim 10^{-2}}_{saddle}/{O(1)-O(10)}_{max}$\\
2   & 3 & $\sim 10^{-3}$ \\
0   & 1 & $ O(1) to O(10) $ \\
\hline\hline
\end{tabular}
\end{center}
\caption{$N_e$ for different choices of the integers $m,n$}
\end{table}

\newpage
\begin{figure}
\epsfxsize=6in
\epsfysize=8.5in
\epsffile{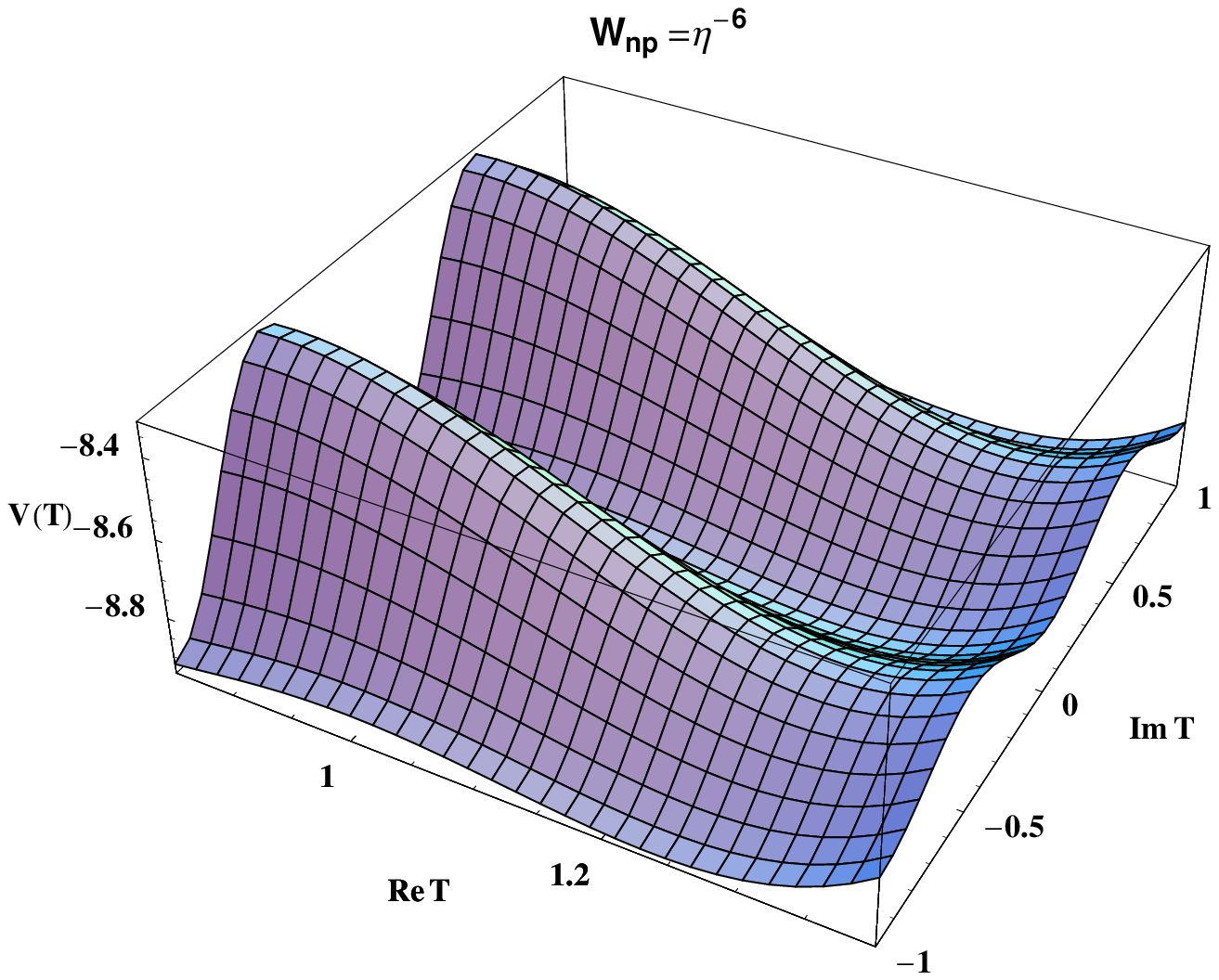}
\caption{$V_{eff}$ for $m=n=0$.}
\end{figure}

\newpage
\begin{figure}
\epsfxsize=6.3in
\epsfysize=8.2in
\epsffile{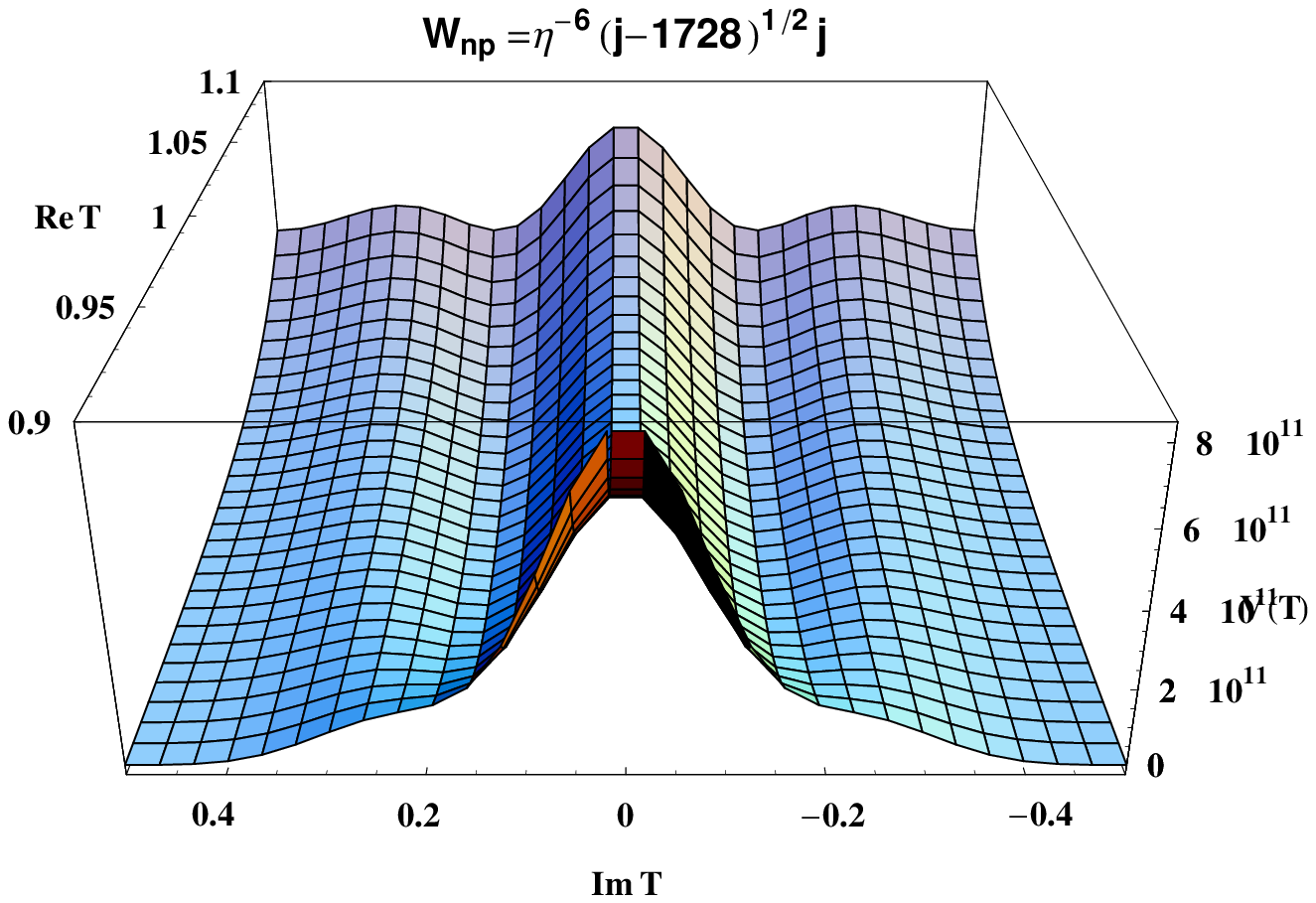}
\caption{$V_{eff}$ for $m=1,n=3$.}
\end{figure}

\newpage
\begin{figure}
\epsfxsize=6.2in
\epsfysize=8.2in
\epsffile{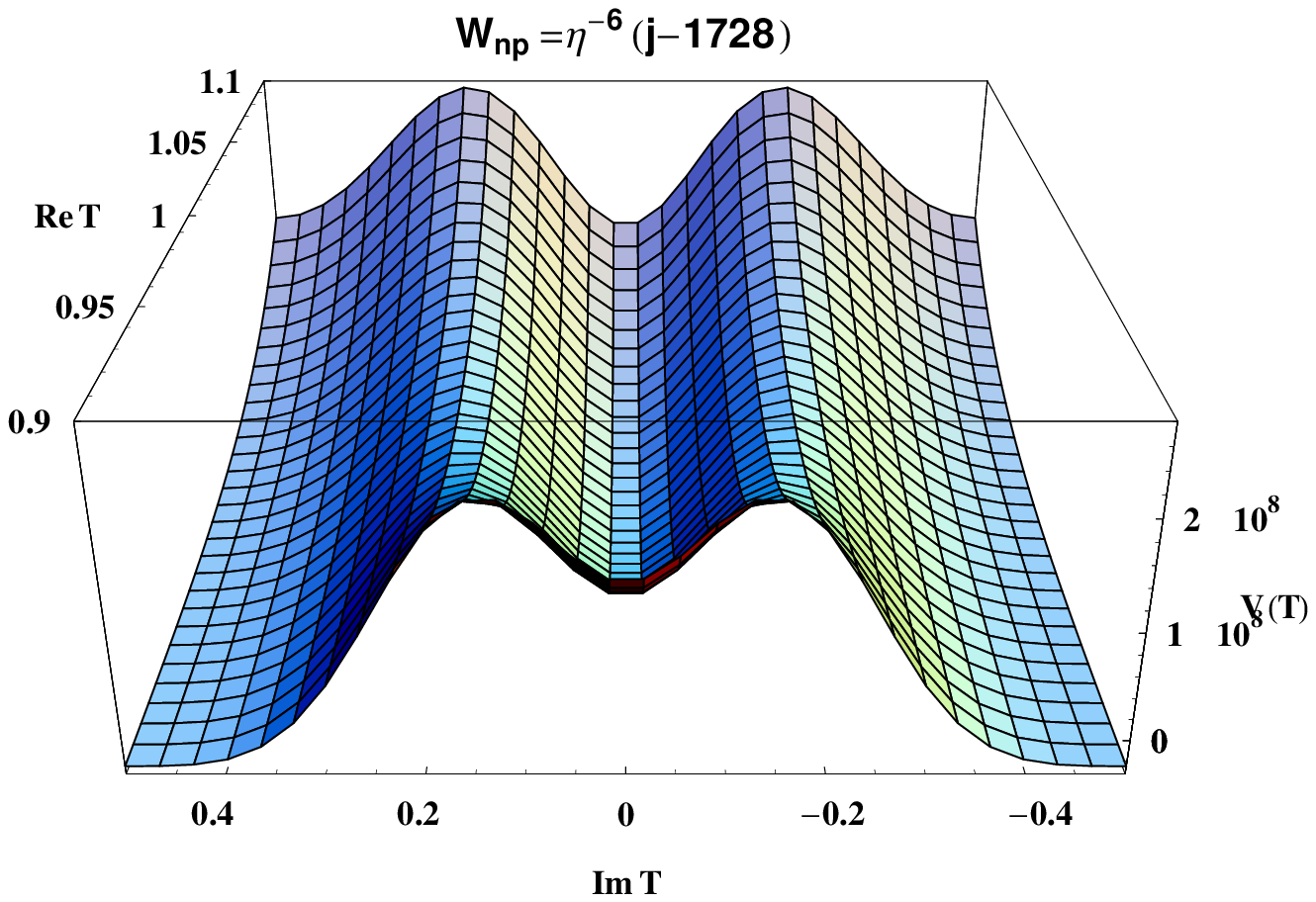}
\caption{$V_{eff}$ for $m=2,n=0$}
\end{figure}

\newpage
\begin{figure}
\epsfxsize=6in
\epsfysize=8.2in
\epsffile{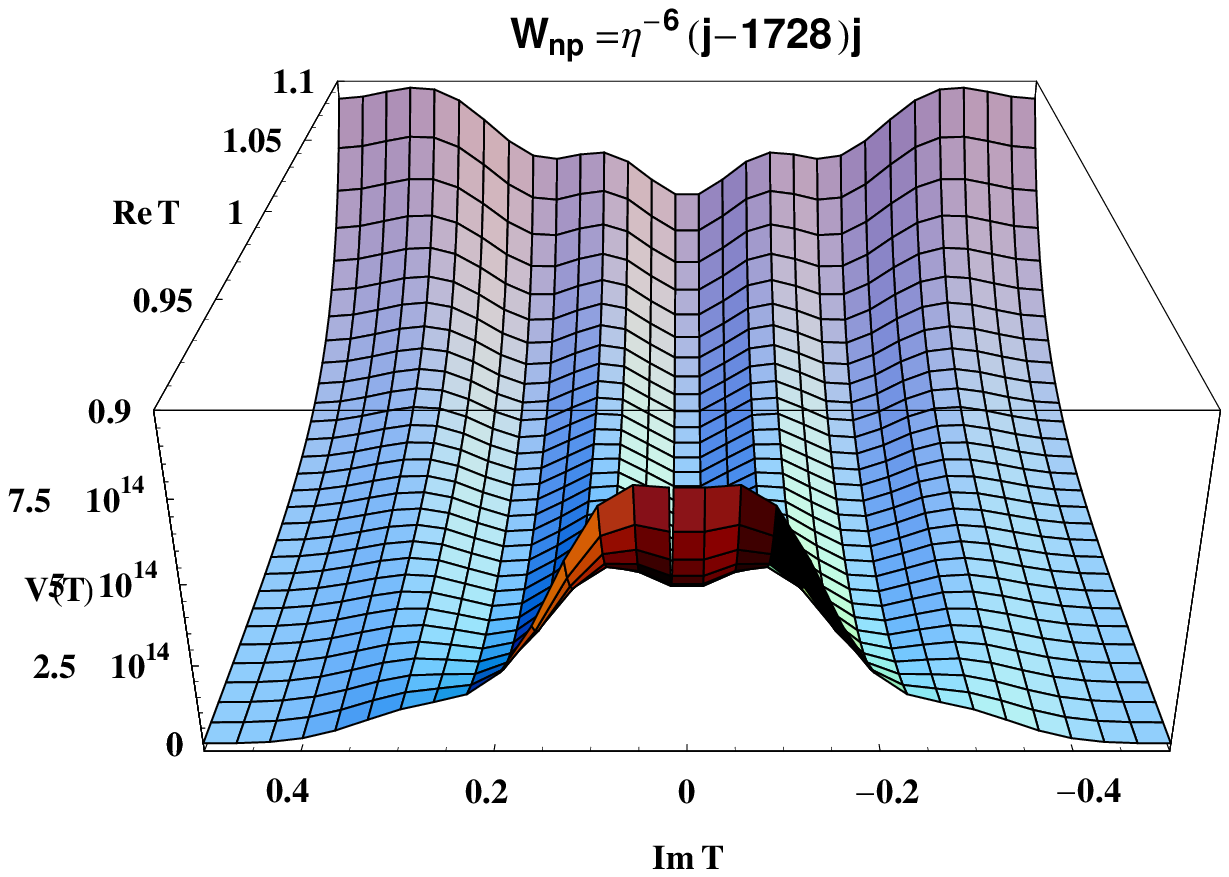}
\caption{$V_{eff}$ for $m=2,n=3$.}
\end{figure}

%%%%%\newpage
%%\begin{figure}
%%\epsfxsize=6in
%%\epsfysize=8.5in
%%\epsffile{modunit.eps}
%%\caption{Potential...}
%%\end{figure}

\newpage
\begin{figure}
\epsfxsize=6in
\epsfysize=8.5in
\epsffile{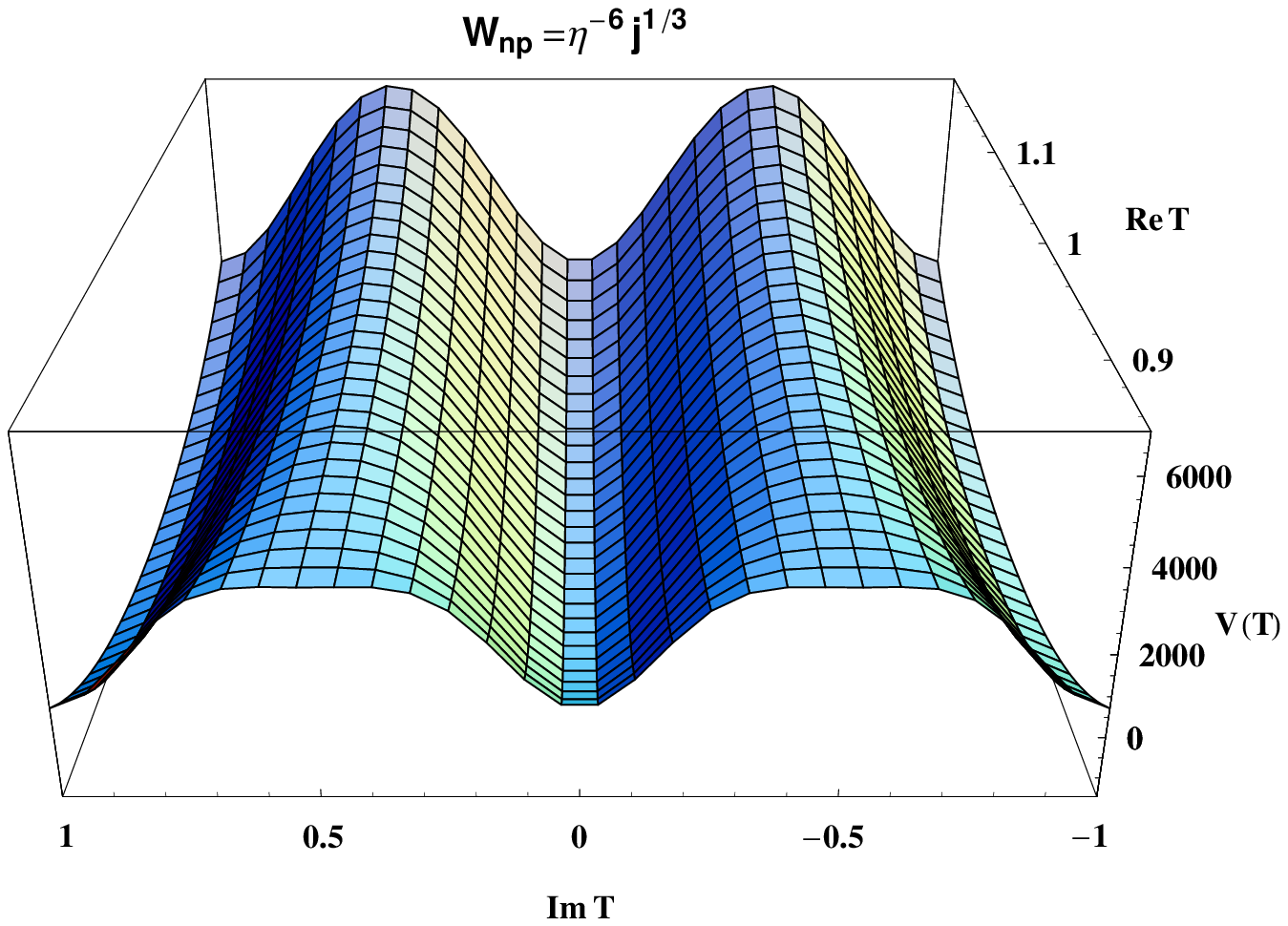}
\caption{$V_{eff}$ for $m=0,n=1$}
\end{figure}

\end{document}